\documentclass{ws-procs11x85}
\usepackage{multicol,float} 

\begin{document}

\title{Models for gamma-ray production 
in low-mass microquasars}

\author{G. S. Vila$^\ddagger$}

\address{Instituto Argentino de Radioastronom\'ia (IAR-CONICET),\\
C.C.5, (1894) Villa Elisa, Buenos Aires, Argentina\\
$^\ddagger$E-mail: gvila@iar-conicet.gov.ar}

\author{G. E. Romero$^{*, **}$}

\address{Instituto Argentino de Radioastronom\'ia (IAR-CONICET),\\
C.C.5, (1894) Villa Elisa, Buenos Aires, Argentina\\
$^{*}$E-mail: romero@iar-conicet.gov.ar\\
Facultad de Ciencias Astron\'omicas y Geofísicas, UNLP,\\
Paseo del Bosque s/n, 1900 La Plata, Buenos Aires, Argentina\\
$^{**}$E-mail: romero@fcaglp.unlp.edu.ar\\}

\begin{abstract}
Unlike high-mass gamma-ray binaries, low-mass microquasars lack external sources of radiation and matter that could produce high-energy emission through interactions with relativistic particles. In this work we consider the synchrotron emission of protons and leptons that populate the jet of a low-mass microquasar. In our model photohadronic and inverse Compton (IC) interactions with synchrotron photons produced by both protons and leptons result in a high-energy tail of the spectrum. We also estimate the contribution from secondary pairs injected through photopair production. The high-energy emission is dominated by radiation of hadronic origin, so we can call these objects \lq proton microquasars\rq.
\end{abstract}


\bodymatter

\begin{multicols}{2}

\section{Introduction}
Microquasars are accreting binary systems that accelerate particles up to relativistic energies, as demonstrated by the detection of synchrotron radiation from radio up to X-ray energies \cite{mir92, cor02}. In microquasars with a high-mass donor star, the relativistic jets ejected from the surroundings of the compact object must traverse photon and matter fields produced by the star. This results in interactions that can produce high-energy gamma rays, as discussed by several authors \cite{rom03, br06}. In fact, three high-mass microquasars have already been detected as gamma ray emitters \cite{aha05, alb06, alb07}. Low-mass microquasars, on the contrary,  have not been detected so far. High-energy emission from these objects has been modeled in Refs. \refcite{br06}, \refcite{aa99}, \refcite{mar01} and \refcite{gren05}. These authors have focused only on leptonic processes. Here we present a new family of models for the non-thermal emission of low-mass microquasars which include several improvements. Among other things, we consider an hadronic content in the jets, proton synchrotron radiation, photomeson and photopair production, leptonic processes for primary and secondary particles, etc. Relativistic particle distributions are calculated in  self-consistent way, solving the transport equation under reasonable assumptions. Our models predict that low-mass microquasars might be detectable by instruments like GLAST and Cherenkov telescopes like VERITAS, HESS and MAGIC, and other future extensions of these arrays.

\section{Model}

We consider an inhomogeneous jet of conical geometry perpendicular to the orbital plane of the binary. The jet is launched at a distance $z_0$ from the compact object and has an initial radius $r_0=0.1\,z_0$. The outflow carries a significant fraction of the accretion power,\cite{fb, kord} $L_{\rm{jet}}= q_{\rm{jet}}\,L_{\rm{accr}}$. We further assume that a fraction $L_{\rm{rel}}=\xi L_{\rm{jet}}$ is in the form of relativistic protons and leptons, $L_{\rm{rel}}=L_p+L_e$. We relate the energy budget of both species as $L_p=aL_e$. Particles are accelerated by diffusive shock acceleration leading to a power-law injection 

\begin{equation}
Q\left(E,z\right)=\frac{Q_0}{z}E^{-\alpha} \qquad [Q]=\rm{erg}^{-1}\rm{s}^{-1}\rm{cm}^{-3}.
\end{equation}

\noindent We consider a canonical value $\alpha=2.2$ for the spectral index, and we assume that the injection rate decreases with the distance to the compact object. Acceleration of particles takes place in a compact region centered at $z_{\rm{acc}}$, with a thickness $\Delta z=5 r\left(z_{\rm{acc}}\right)$. The efficiency of the acceleration mechanism is characterized by the parameter $\eta$, in such a way that the acceleration rate is 

\begin{equation}
t_{\rm{acc}}^{-1}=\eta c e B E^{-1}.
\end{equation}

\noindent We assume an efficient accelerator with $\eta=0.1$. The magnetic field in the jet decreases as the jet expands as $B=B_0(z_0/z)$. Its value at $z_0$ is determined by equipartition between magnetic and kinetic energy densities, $U_B(z_0)=U_{\rm{jet}}^{\rm{kin}}(z_0)$. This yields $B_0\sim10^7$ G and makes synchrotron losses the dominant channel of radiative particle cooling. For a particle of mass $m$ and energy $E$, the synchrotron cooling rate is given by

\begin{equation}
t_{\rm{synchr}}^{-1}=\frac{4}{3}\left(\frac{m_e}{m}\right)^3\frac{c\sigma_TU_B}{m_ec^2}\frac{E}{mc^2}.
\end{equation}

\noindent Particle distributions in the steady state $N_{p,e}(E_{p,e},z)$ (erg$^{-1}$cm$^{-3}$) are calculated solving a transport equation that takes into account particle injection, cooling and particle escape from the acceleration region, 

\begin{equation}
\frac{\partial}{\partial E}\left[\left.\frac{dE}{dt}\right|_{\rm{loss}}N\left(E,z\right)\right]+\frac{N\left(E,z\right)}{t_{\rm{esc}}}=Q\left(E,z\right),
\label{transp_eq}
\end{equation}

\smallskip
                    
\noindent where $\left.dE/dt\right|_{\rm{loss}}$ is the sum of all radiative and mechanical losses for each type of particle.                     
                            
\begin{figure*}[!t]
\begin{center}
\psfig{file=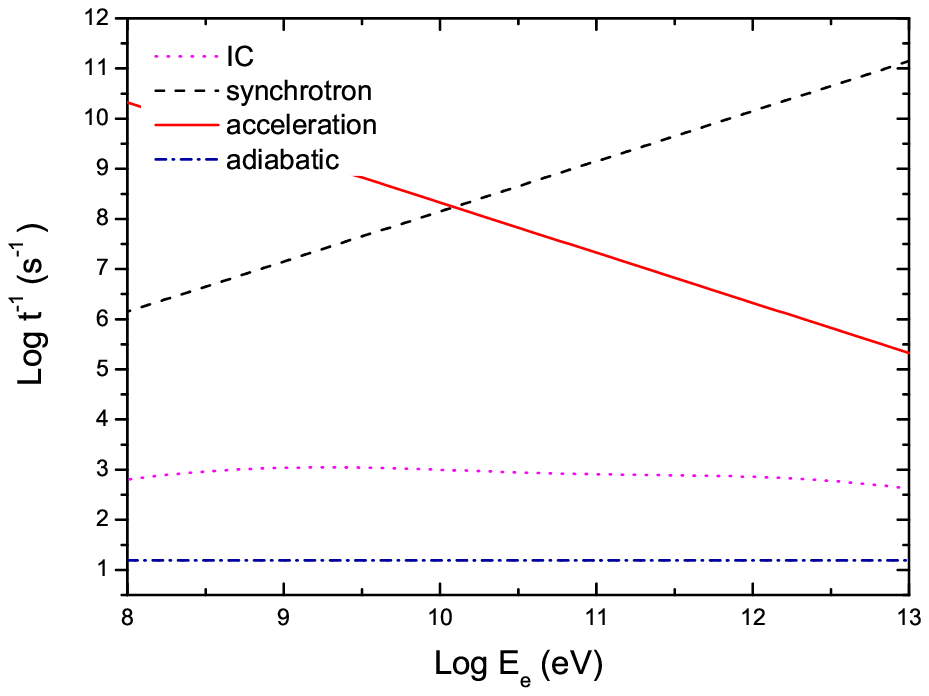,trim=0 17 0 0, clip, width=0.49\textwidth}
\psfig{file=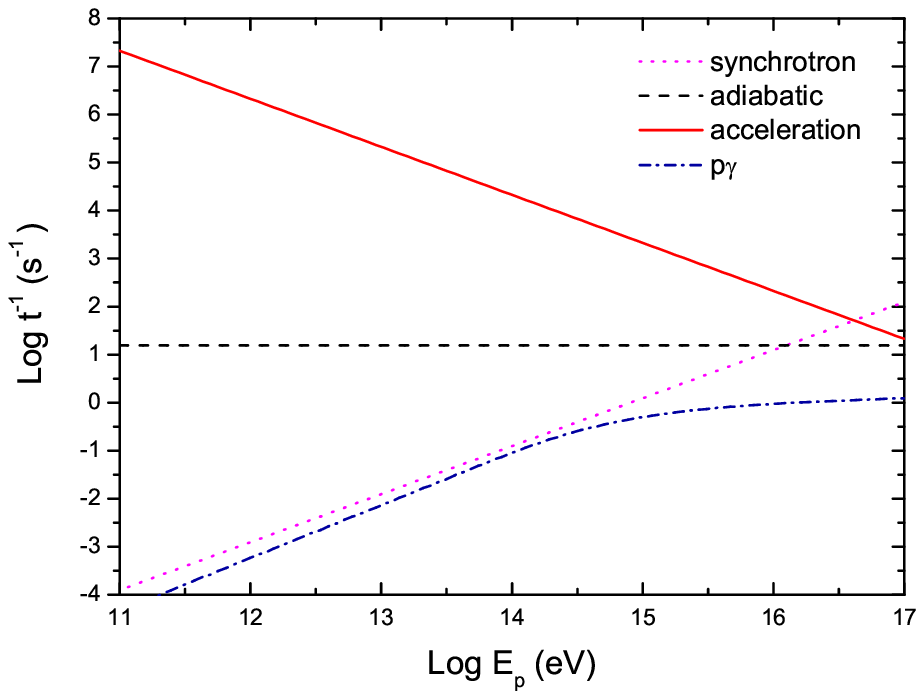,trim=0 17 0 0, clip,width=0.48\textwidth}
\end{center}
\caption{Acceleration and cooling rates at $z_{\rm{acc}}$ for electrons (left panel) and protons (right panel), for  $a=1$. In the case of protons, the $p\gamma$ cooling rate is the sum of the photopair and photomeson production contributions. The dominant channel of energy loss for electrons is synchrotron radiation. For protons, adiabatic (low energies) and synchrotron (high energies) losses are the main cooling mechanisms. 
}
\label{aba:fig2}
\end{figure*}

\noindent We estimate the escape time as $t_{\rm{esc}}= z_{\rm{acc}}/v_{\rm{jet}}$, where $v_{\rm{jet}}\sim0.7\,c$ corresponds to a bulk Lorentz factor $\Gamma_{\rm{jet}}=1.5$, typical of a jet in the low-hard state of the microquasars\cite{Fen2}. The proton spectrum goes as  $N_p\propto E_p^{-2.2}$ for all $E_p$. In the case of electrons, instead, synchrotron losses introduce a steepening that leads to $N_e\propto E_e^{-3.2}$ at all energies before the cutoff. The acceleration and cooling rates for both protons and electrons at the acceleration region are shown in Figure \ref{aba:fig2}. The maximum energy of all particles is determined by the synchrotron cooling rate. In Table \ref{aba:tab1} we show some representative values of the relevant parameters.

\begin{table}[H]
\tbl{Values of the various parameters characterizing the jet and the distributions of relativistic particles that were used in the calculations.}
{\begin{tabular}{@{}ll@{}}
\toprule
Parameter & Value \\ \colrule
$L_{\rm{accr}}$: accretion power & $1.7\times10^{39}$ erg s$^{-1}$$^{(1)}$\\
$q_{\rm{jet}}$: coupling constant accretion-jet power & $0.1$ \\ 
$z_0$: jet's launching point & $2\times10^8$ cm$^{(2)}$ \\
$z_{\rm{acc}}$: location of the acceleration region& $10^{9}$ cm \\
$\Delta z$: thickness of the acceleration region& $5 r(z_{\rm{acc}})$ \\
$\Gamma_{\rm{jet}}$: jet's bulk Lorentz factor & $1.5^{(1)}$ \\
$\theta$: viewing angle & $30^\circ$ \\
$\xi$: ratio $L_{\rm{rel}}/L_{\rm{jet}}$ & $0.1,\; 10^{-3}$ \\
$a$: ratio $L_p/L_e$ & $1, \;100,\; 10^3$ \\
$\eta$: acceleration efficiency & $0.1$ \\
$B_0$: magnetic field at $z_0$ & $2\times10^7$ G\\
$E^{\rm{min}}_{p,e}$: minimum particle energy & $100\,m_{p,e}c^2$ \\
$E^{\rm{max}}_{p,e}$: maximum particle energy $p,e^-$ & $10^{10},\,7\times10^{16}$ eV$^{(3)}$ \\ \botrule
\multicolumn{2}{l}{
$^{(1)}$ Typical value for the low-mass microquasar XTE J1118+480\cite{mark00}}\\ 
\multicolumn{2}{l}{
$^{(2)}$ $50 R_{\rm{Schw}}$ for a $8\,M_{\odot}$ black hole as XTE J1118+480\cite{gel06} }  \\
\multicolumn{2}{l}{$^{(3)}$ These are the maximum values obtained along the jet.}  \\[0.01cm]
\end{tabular}}
\label{aba:tab1}
\end{table}

\section{Results}

In addition to synchrotron radiation, we consider other two photon production mechanisms:  IC and proton-photon colissions ($p\gamma$) on the synchrotron radiation fields of both protons and leptons. The later process has two main branches: photomeson production 

\begin{equation*}
	p+\gamma\rightarrow p+a\pi^0+b\left(\pi^++\pi^-\right)
\end{equation*}

\bigskip

\noindent and photopair production  

\begin{equation*}
	p+\gamma\rightarrow p+e^++e^-
\end{equation*}

\bigskip

\noindent The spectrum from the decay of  $\pi^0\rightarrow2\gamma$ is estimated following Ref. \refcite{ad03}. For the injection of electron/positron pairs we use the formulae given by Refs. \refcite{ch92} and \refcite{mast05}. The IC spectra are calculated in the local approximation of Ref. \refcite{ghis85}, whereas for synchrotron radiation we use the classical expressions, as given, for example, by Ref. \refcite{bg70}.  All the calculations are performed in the jet co-moving reference frame and then transformed to the observer frame using the appropriate Doppler factor. We fix $\theta=30^\circ$ as the viewing angle for numerical estimates. Since the jet is only mildly relativistic, none of our results depends strongly on this value as long as it is not very close to zero.

\begin{figure*}[!t]
\begin{center}
\psfig{file=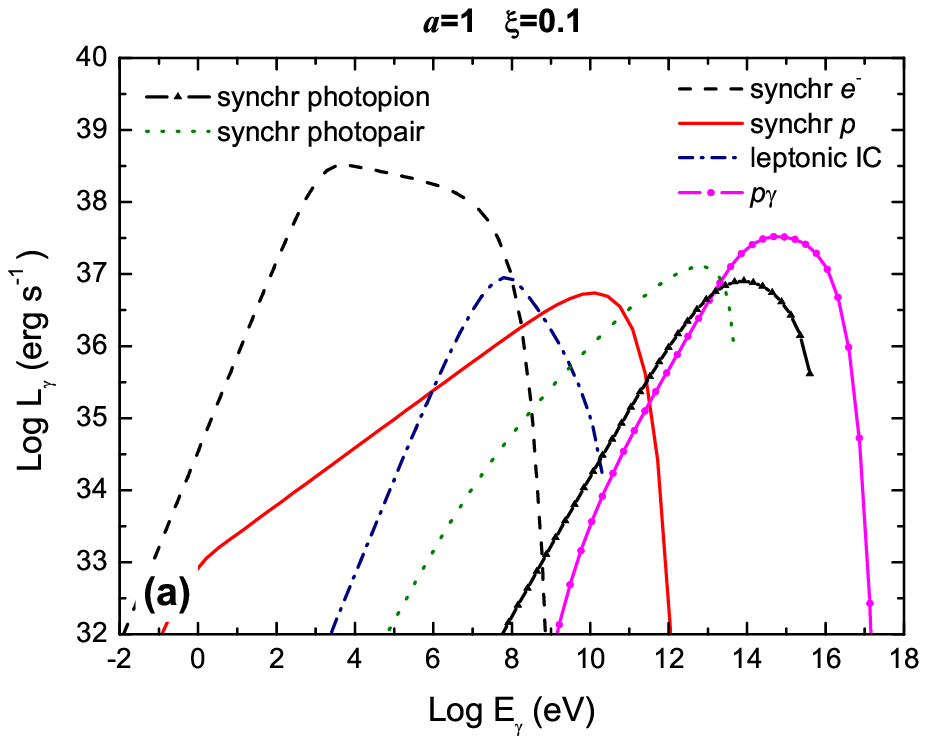,trim=0 17 0 5, clip, width=0.45\textwidth}
\psfig{file=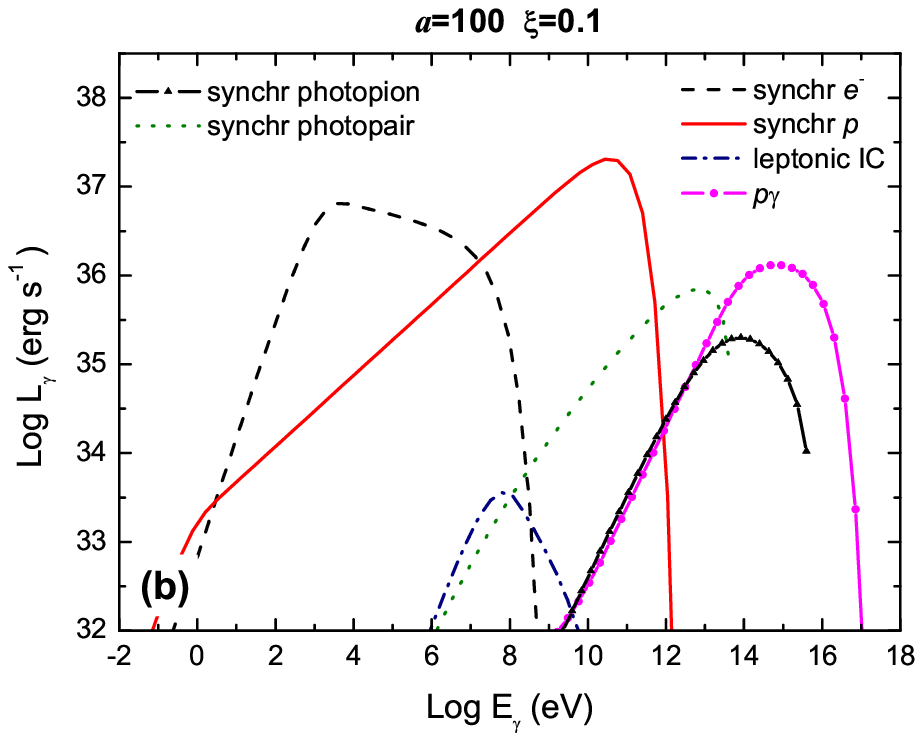,trim=0 17 0 5, clip, width=0.46\textwidth}
\psfig{file=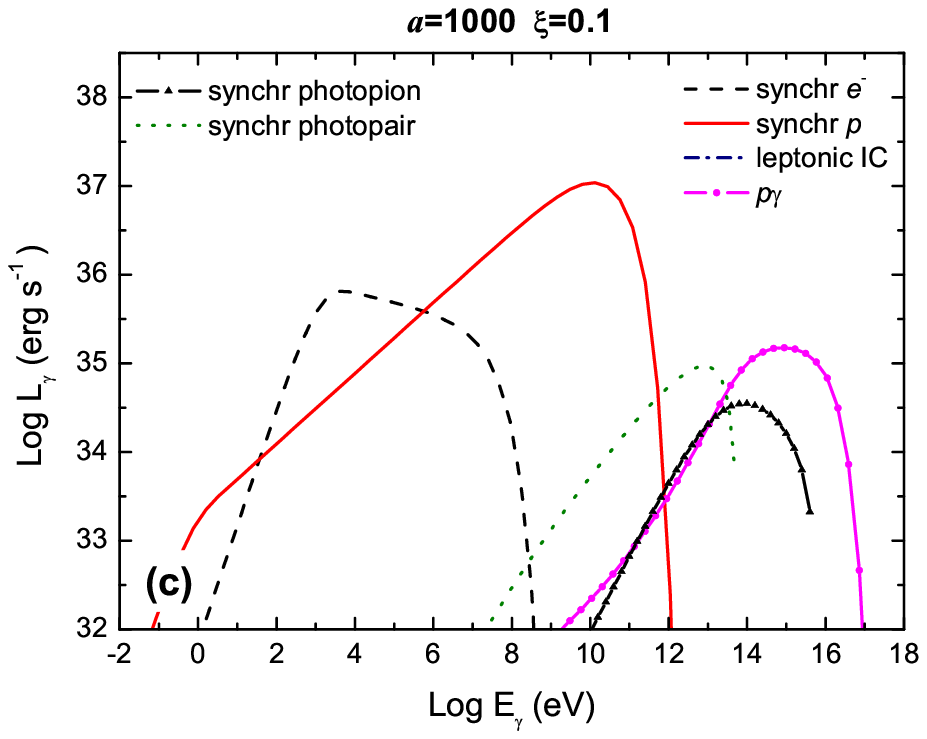,trim=0 17 0 5, clip, width=0.45\textwidth}
\psfig{file=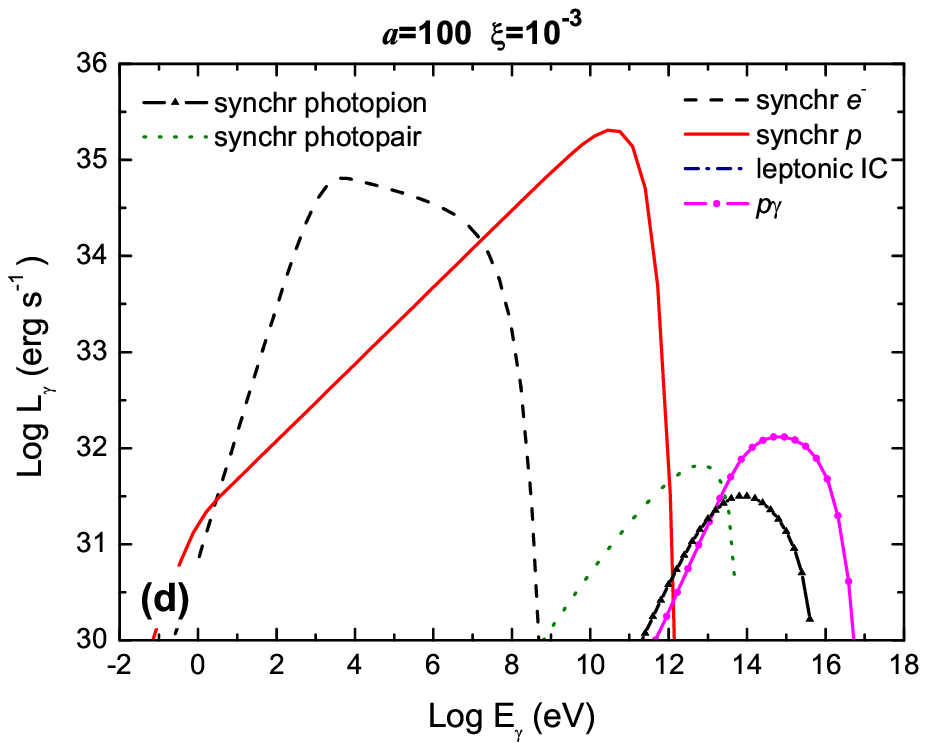,trim=0 17 0 5, clip, width=0.45\textwidth}
\end{center}
\caption{Production spectra obtained for different values of jet content of relativistic particles $\xi$, and proton-to-lepton luminosity ratio $a$. 
}
\label{aba:fig1}
\end{figure*}

Figure \ref{aba:fig1} (b), (c) and (d) correspond to a proton-dominated jet. The peak of the spectral energy distribution in the three cases is due to the proton synchrotron radiation. At energies beyond 10 TeV, the dominant contribution comes from the decay of neutral pions produced through photomeson mechanism. In the energy range between 1 TeV and 100 TeV, the synchrotron radiation from photopairs is systematically more significant than that generated by secondary pairs from charged pion decays. 

Case (a) corresponds to equipartition between protons and leptons ($a=1$). As expected, the lepton synchrotron luminosity rises. A denser photon field results in an enhanced (although relatively small)  leptonic inverse Compton (IC) emission, but an appreciable $p\gamma$ luminosity at very high energies, characterized by a hard spectrum. In case (d) the relativistic particle content is smaller,  $\xi=L_{\rm{rel}}/L_{\rm{jet}}=10^{-3}$. This reduces synchrotron, IC and $p\gamma$ luminosities by one and two orders of magnitude, respectively. 

Concerning the radio emission, our models, which have a high minimum injection energy for electrons, are not intended to reproduce the observable radio spectrum of systems like XTE J1118+480 ($L_{\rm{radio}}\sim 10^{28-29}$ erg s$^{-1}$ \cite{fen01b}). However, models with $E^{\rm{min}}\sim 2m_ec^2$ can yield the expected luminosity and spectrum in this band\cite{rom08}. 

\section{Conclusions}

Low-mass microquasars  with steady jets can be significant sources of high-energy radiation if they have an important hadronic content. These objects can be found out of the galactic plane, because they have large proper motions\cite{mir01}. They can also be sources of neutrinos through charged pion decays from $p\gamma$ interactions, as already noticed by Ref. \refcite{lw01}. These \lq proton microquasars\rq\ might be detected in the near future by GLAST and ground-based Cherenkov telescope arrays, since gamma-ray luminosities in the range $10^{33}-10^{36}$ erg s$^{-1}$ between 1 GeV and 10 TeV can be expected under a variety of conditions. Moreover, some of the variable, unidentified sources detected by EGRET around the galactic center at  $E_\gamma>100$ MeV could be \lq proton microquasars\rq\ ejected from the halo or the plane of the Galaxy.   

\section*{Acknowledgments}

We thank an anonymous referee for constructive comments. This work has been supported by grants PIP 5375 (CONICET) and PICT 03-13291 BID 1728/OC-AR (ANPCyT). GER acknowledges support by the Ministerio de Educaci\'on y Ciencia (Spain) under grant AYA 2007-68034-C03-01, FEDER funds.

\end{multicols}
\end{document}